# Stochastic Nonlinear Electrical Characteristics of Graphene

Young Jun Shin,[1] Kalon Gopinadhan,[1,2] Kulothungasagaran Narayanapillai,[1] Alan Kalitsov,[1] Charanjit S. Bhatia,[1] and Hyunsoo Yang[1,2,a)]

[1]*Department of Electrical and Computer Engineering and NUSNNI-Nanocore, National University of Singapore, 117576, Singapore*

[2]*Graphene Research Centre, Faculty of Science, National University of Singapore, 117546, Singapore*

A stochastic nonlinear electrical characteristic of graphene is reported. Abrupt current changes are observed from voltage sweeps between the source and drain with an on/off ratio up to $10^3$. It is found that graphene channel experience the topological change. Active radicals in an uneven graphene channel cause local changes of electrostatic potential. Simulation results based on the self-trapped electron and hole mechanism account well for the experimental data. Our findings illustrate an important issue of reliable electron transports and help for the understanding of transport properties in graphene devices.

[a)] e-mail address: eleyang@nus.edu.sg

Graphene, two-dimensional carbon crystal, has attracted enormous attention because of its very unique and superior physical properties.[1-4] Many interesting physical phenomena of Dirac fermions in monolayer graphene have been explored in the laboratories. Simultaneously, graphene has been investigated for commercialization purpose by taking advantage of its outstanding electrical and optical properties. Graphene is a very promising material for high frequency devices because of its superior mobility at room temperature.[5] Most importantly, graphene is one of the best candidates for the transparent contact material. Indium Tin Oxide (ITO) has been dominantly used as transparent contacts for display and photovoltaic devices. However, because of high cost and limited supply of indium, many alternative materials have been widely explored to rule over the post-ITO era. Chemical vapor deposition (CVD) based growth and chemical doping render graphene competitive over ITO in terms of large scalability and the sheet resistance.[6, 7] Since graphene is very robust and flexible, graphene becomes highly attractive than brittle ITO for flexible display applications.[8, 9]

Consequently, major electronics companies have already started commercialization activities including a pilot line for large scale graphene growth. However, is graphene really qualified for the real applications? Graphene is known as a good candidate for chemical or gas sensors due to its large surface to volume ratio.[10] It has been demonstrated that graphene is capable of detecting individual gas molecules.[11] Is graphene suitable for stable electrical contacts, if graphene is a very sensitive material? It has been studied that the insulating state of bi-layer graphene can be converted to the metallic state by exposing pristine bi-layer graphene to atmosphere.[12] Maintaining stable electrical properties under various circumstances, especially for long term use with frequent sweeps of the operational bias voltage, is very crucial for graphene to be integrated into display electronics as transparent contacts. Therefore, the stability and reliability of graphene should be investigated more carefully and a deep understanding is necessary.

In this study, we present a stochastic nonlinear electrical characteristic of graphene. Abrupt current changes are observed from voltage sweeps between the source and drain with an on/off ratio up to $10^3$. In order to understand the origin of nonlinear behavior, electrical transport measurements are conducted under controlled environments and microscopic characterization has been done after observing the random transitions. The simulation results based on the self-trapped electron and hole mechanism account well for the experimental data. Our findings illustrate an important issue of reliable electron transports and help for the understanding of transport properties in graphene devices.

Two-terminal graphene devices with a back gate are fabricated on top of highly p-doped Si substrates covered by a 300 nm thick $SiO_2$ layer as shown in the inset of Fig 1(a). Mechanically exfoliated single and multi layer graphene is utilized in this study and we are able to observe the nonlinear switching effect regardless of the number of graphene layer. The quality of graphene is checked with optical microscopy, AFM, and Raman spectroscopy. For the device fabrication, optical and e-beam lithography are used for electrode patterns and Cr/Au (5 nm/120 nm) is deposited by a thermal evaporator, followed by lift-off. Further details on the sample preparation, identification, and fabrication can be found elsewhere.[13]

Figure 1(a) shows many experimental *I-V* switching curves from a single device. The bias voltage is swept from 0 to 3 V and back to 0 V between source and drain with zero back gate bias, repeatedly under ambient conditions. The random transitions between a low-resistive, metallic state and a high-resistive, insulating state are observed. The three most representative switching phases from *I-V* traces in Fig. 1(a) are plotted separately in Fig. 1(b-d). For example, an ON-OFF transition in Fig. 1(c) represents that the device starts from an ON state when the bias voltage increases from zero voltage, and it ends with an OFF state when the voltage sweeps back to zero bias. The resistance of two-terminal graphene devices is ~a few kΩ when it is in metallic (ON) state, whereas the resistance is higher than a few MΩ in the insulating (OFF) state.

The transition can be more clearly seen, when the *I-V* data are plotted in a logarithmic scale as shown in Fig. 2(a). The difference of the current between ON and OFF states is more than three orders of magnitude. When the resistance of the device is plotted as a function of measurement time in Fig. 2(b), there are random transitions among three switching phases and a switching histogram of phases is shown in Fig. 2(c). From the statistical study, the ON-OFF (or OFF-ON) phase is found to be less frequent than either the ON-ON or OFF-OFF phase. Fig. 2(d) shows the resistance as a function of back gate bias at two different states; ON and OFF. The charge carrier density of graphene is $4\times10^{12}$ $cm^{-2}$ at zero back gate bias in a metallic state. The junction resistance at a metallic state follows a well-defined typical p-type Dirac curve (black line). On the other hand, a negligible change due to the back gate voltage is found in an insulating state (red line). A Dirac curve is observed again when the phase is changed back from insulating to metallic by voltage sweep (black dots), demonstrating reliable transitions between two ON and OFF states.

It is well known that graphene can be easily influenced by external doping sources such as absorbed molecules.[1] An opposite sequence of current hysteresis originated from attaching and detaching active radicals from the air has been reported.[14] Tunable metal-insulator transitions (MIT) in bilayer graphene caused by water vapors has been also found.[12] Electrochemical reactions caused by active radicals can be a key to elucidate the origin of the observed stochastic nonlinear effect. In order to understand the underlying mechanism, many repeated *I-V* sweeps are conducted in a vacuum chamber (< $1\times10^{-7}$ Torr) for the samples which show random transitions at ambient conditions. All the active radicals are first detached from the channel by applying high currents through the graphene channel under a vacuum because of the thermal energy generated by current annealing. After removing all doping sources, the nonlinear random transitions have never been observed with more than 20 devices measured in a vacuum as shown in Fig. 3(a). When the device is exposed to air again, the random transitions are reinstated. From this result, it is clear that the observed phenomenon is strongly correlated to electrochemical reactions caused by active radicals which is attached and detached to the graphene channel from air. Similar phenomenon with a top gate structure has been reported observing reversible bipolar switching by applying electrochemical modification.[15] In this case, hydrogen and hydroxyl, catalytically produced in the silicon oxide top gate, work as chemical doping sources and devices did not exhibit the switching anymore under high vacuum conditions similar to our case.

In order to investigate any mechanical deformation or topological changes associated with the random transitions, material characterization techniques such as atomic force microscopy (AFM) and scanning electron microscopy (SEM) are employed to image the graphene channels. All imaging have been done after the transport measurements, since the characterization may damage or change the graphene property. As can be seen from Fig. 3(b), the SEM image of the graphene channel is not flat and a topological change of graphene is observed. The mechanically deformed part which may induce a local strain due to an upheaval structure is clearly visible by the AFM measurement in Fig. 3(c). The current density between two terminals is usually higher than a few times $10^8$ A/cm$^2$ at 3 V, when the device is in a metallic state. In this regime, it is reasonable to have electromigration and large Joule heating effects.[16] Heat dissipation more than a few mW may cause the topological change of graphene. We have investigated more than 300 exfoliated graphene and cannot observe any upheaval or local strain from the pristine exfoliated graphene. We can observe the deformed graphene channel only after measuring stochastic transitions, indicating that the deformed graphene is responsible for the random transitions. In order to support our assumption, we sweep the voltage in a small range (< 0.1 V) for more than 50 devices. Only typical ohmic *I-V* characteristics are measured rather than the nonlinear behavior, and no deformed graphene channel is found after the transport measurements. This is in line with previous studies, in which metal-semiconductor transition (MST) and metal-insulator transition has been observed from the two-terminal carbon nanotube (CNT) devices exposed to electron beam irradiation.[17, 18] Inhomogeneous electric fields generated from the trapped charges in $SiO_2$ are proposed to be the origin in this case.

We conclude that the observed random transition of graphene is attributed to unevenly attached active radicals, similarly working as the trapped charges to create inhomogenuous fields, in a non uniform graphene channel. Since our graphene channel is inhomogeneous, unevenly attached active radicals in graphene locally change the electrostatic potential. We have done transport simulations considering self-trapped electrons and holes in the upheaval graphene channel caused by electrochemical reactions. We consider the tight binding Hamiltonian of graphene cluster taking into account electron-electron interactions

$$H_G = -t\sum_{i,j,\sigma}(a^+_{i\sigma}b_{j\sigma} + b^+_{i\sigma}a_{j\sigma}) + U\sum_i n_i^{\uparrow} n_i^{\downarrow} \,, \tag{1}$$

where $t$ is the spin-independent effective hopping integral between the nearest-neighbour carbon atoms, $U$ is the electron-electron interaction parameter, $n^{\sigma}{}_i$ is the operator of number of electrons with spin $\sigma$, $a^+{}_{i\sigma}$ ($b^+{}_{i\sigma}$) and $a_{i\sigma}$ ($b_{i\sigma}$) are the creation and annihilation operators of the conduction electron with spin $\sigma$ on site $i$ on graphene sublattice A (B). We apply the Hartree-Fock approximation for the second term of eq.(1) and rewrite the Hamiltonian of the graphene cluster

$$H_G = -t\sum_{i,j,\sigma}(a^+_{i\sigma}b_{j\sigma} + b^+_{i\sigma}a_{j\sigma}) + \frac{U}{2}\sum_{i,\sigma}\left(\langle n_i\rangle - 1\right)(a^+_{i\sigma}a_{i\sigma} + b^+_{i\sigma}b_{i\sigma}) \,, \tag{2}$$

where $<n_i>$ is the average number of electrons on site $i$. For uncharged graphene $<n_i>=1$, while $<n_i>=2(0)$ if an extra electron (or hole) is on site $i$. The graphene cluster is coupled with two non-magnetic leads. The Hamiltonian of the device has the form

$$H = H_G + H_L + H_R + H_{GL} + H_{GR} + H.c.\,, \tag{3}$$

where $H_{L/R}$ is the Hamiltonian of the left/right lead, and the term $H_{GL/GR}$ describes the coupling of the graphene cluster to the left/right lead. We calculate the electric current through the system, when voltage is applied across the devices. Our calculations are based on the non-equilibrium Green functions formalism. The details of the approach can be found elsewhere.[19] First we diagonalize $H_G$ and find the retarded Green function of the uncoupled graphene cluster $g^r$. Next we find the retarded Green function of the coupled system by solving the Dyson equation

$$G^r = g^r + g^r \Sigma_L^r G^r + g^r \Sigma_R^r G^r, \tag{4}$$

where $\Sigma_{L/R}^r$ is the retarded self-energy due to connection of the graphene cluster to the left/right lead. We assume that $\Sigma_{L/R}^r$ is independent of energy. The final expression for the charge current becomes

$$I = \frac{e}{h}\int d\varepsilon [f_L(\varepsilon) - f_R(\varepsilon)] Tr[G^a \Gamma_R G^r \Gamma_L], \tag{5}$$

where $G^a$ is the advanced Green functions of the coupled system, $\Gamma_{L/R} = i(\Sigma_{L/R}^r - \Sigma_{L/R}^a)$ and $f_{L/R}$ is the Fermi-Dirac distribution functions in the left/right lead. We calculate the *I-V* characteristics assuming that the charge distribution on the graphene surface randomly changes, when the applied voltage reaches the threshold value of 3 V. Based on this, we obtain the result of *I-V* traces, which is very similar to the experimental results, as ploted in Fig. 3(d). The inset in Fig. 3(d) shows the resistance as a function of simulation time. There are also random transitions among three switching phases as similar to the experimental data, which can be seen from Fig. 2(b). However, further studies are required for better understanding of the physical mechanism of the random transitions.

It was reported previously that mechanical discontinuity of graphene and graphitic nano-ribbon can be the origin of the switching effect. We scrutinize all the samples which show random transitions with AFM and SEM very carefully and cannot find any mechanical discontinuity across the graphene channels. However, there are mechanical deformations. When the sweep voltage is higher than its saturation value, graphene devices are burnt rather than being discontinuous.[20]

In order to rule out other possibilities as the origin of the observed effect such as the current annealing effect and interface issues between graphene and contacts, the graphene devices are annealed under high vacuum conditions at 500 K for 2 hours. After annealing, a random hysteresis is still observed under ambient conditions. Since the electrical property of graphene is highly dependent on the level of defects, we intentionally introduce large defects into graphene by an oxygen plasma treatment and confirm the level of defect by Raman spectroscopy. Graphene is electrically annealed after the oxygen plasma treatment and electrical transport measurements have been conducted. Although the current hysteresis is found, the hysteresis is not repeatable and the value of the tolerable current is very small (order of several μA).

Frequent breakdowns are occurred during the *I-V* sweeps because of a high electrical field used in the experiments. When the graphene channel becomes a mixed phase structure with *sp*2 and *sp*3 fractions, the tunneling characteristic becomes the main transport mechanism as can be seen in Fig. 4.[20] In our case the samples undergo the following breakdown sequence such that the transport property changes from the ohmic to random transition, and then to tunneling, finally resulting in a complete breakdown (open circuit).

In conclusion, we have reported stochastic transitions between an ohmic like state and an insulator like stage in graphene devices. It is found that the topological change in the graphene channel is involved for the observed behavior. Active radicals with an uneven graphene channel cause a local change of electrostatic potential, and simulations based on the self-trapped electron and hole mechanism can account for the observed data. Understanding electrical transport of graphene at room temperature and at high bias voltages is very important for the interconnect and transparent contact applications. Moreover, further investigations may open up a promising way to engineer graphene memories and logic devices with a high on/off ratio.

## Acknowledgements

The study was supported by the Singapore National Research Foundation under CRP Award No. NRF-CRP 4-2008-06.

Figure Captions

Figure 1. (a) Experimental *I-V* curves of a two-terminal single layer graphene device. The inset in (a) shows a schematic of graphene device. Three most representative switching phases: (b) ON-ON, (c) ON-OFF (or OFF-ON), and (d) OFF-OFF.

Figure 2. (a) Current as a function of the channel bias voltage in a logarithmic scale. (b) Resistance as a function of measurement time. (c) A histogram of three representative phases. (d) Resistance versus back gate voltage ($V_g$) of a device in metallic and insulating phases.

Figure 3. (a) *I-V* curves in vacuum. (b) A scanning electron microscopy image of graphene channel after observing stochastic transitions. (c) An atomic force microscopy image of graphene channel indicated as a red box in (b). The bottom figure is the line scan of the red line. (d) Simulated switching *I-V* curves. The inset in (d) shows the resistance as a function of simulation time.

Figure 4. Resistance as a function of measurement time. The inset shows a typical *I-V* curve in the tunneling regime. The stochastic nonlinear switching behavior has been observed before the tunneling regime.

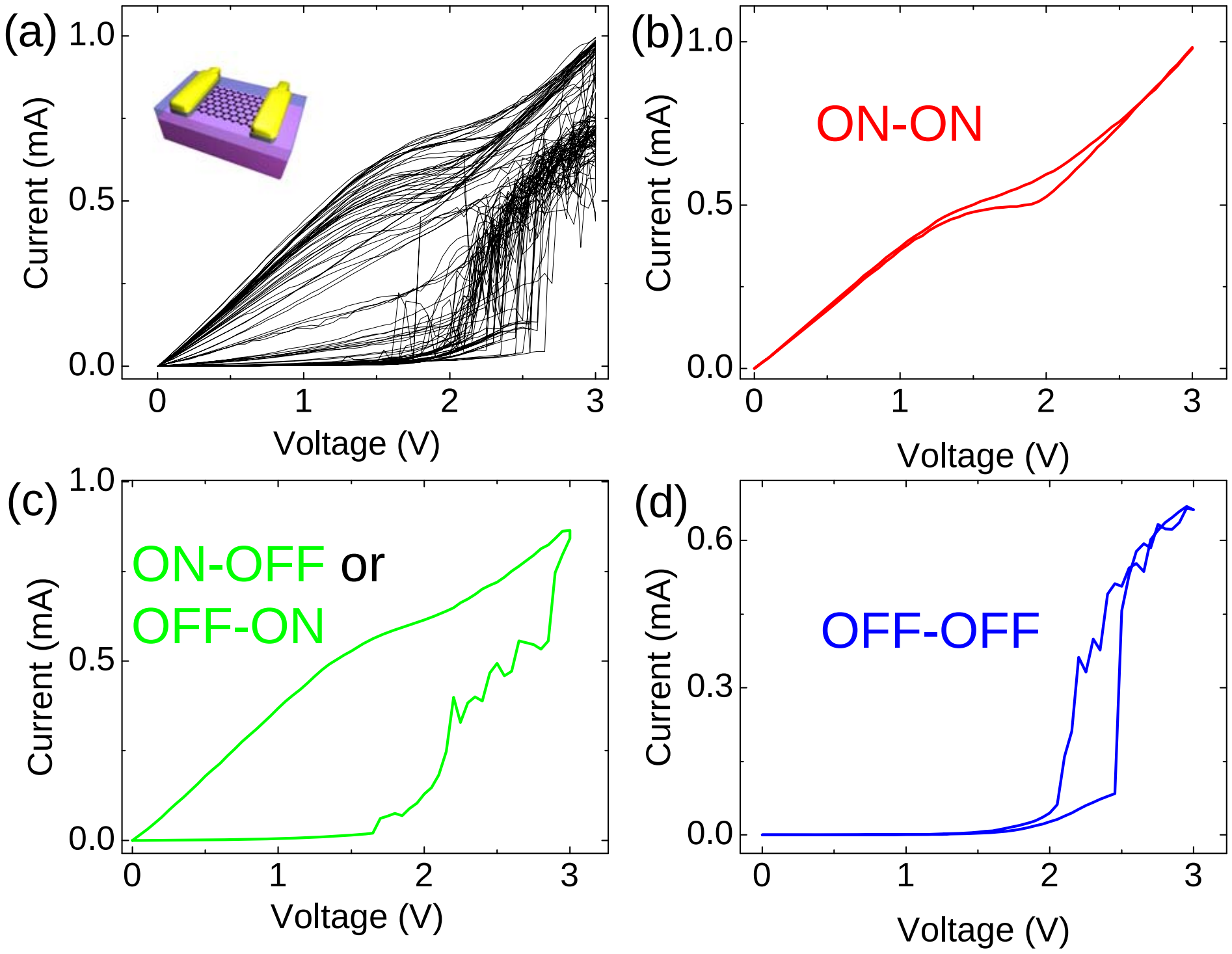

Figure 1.

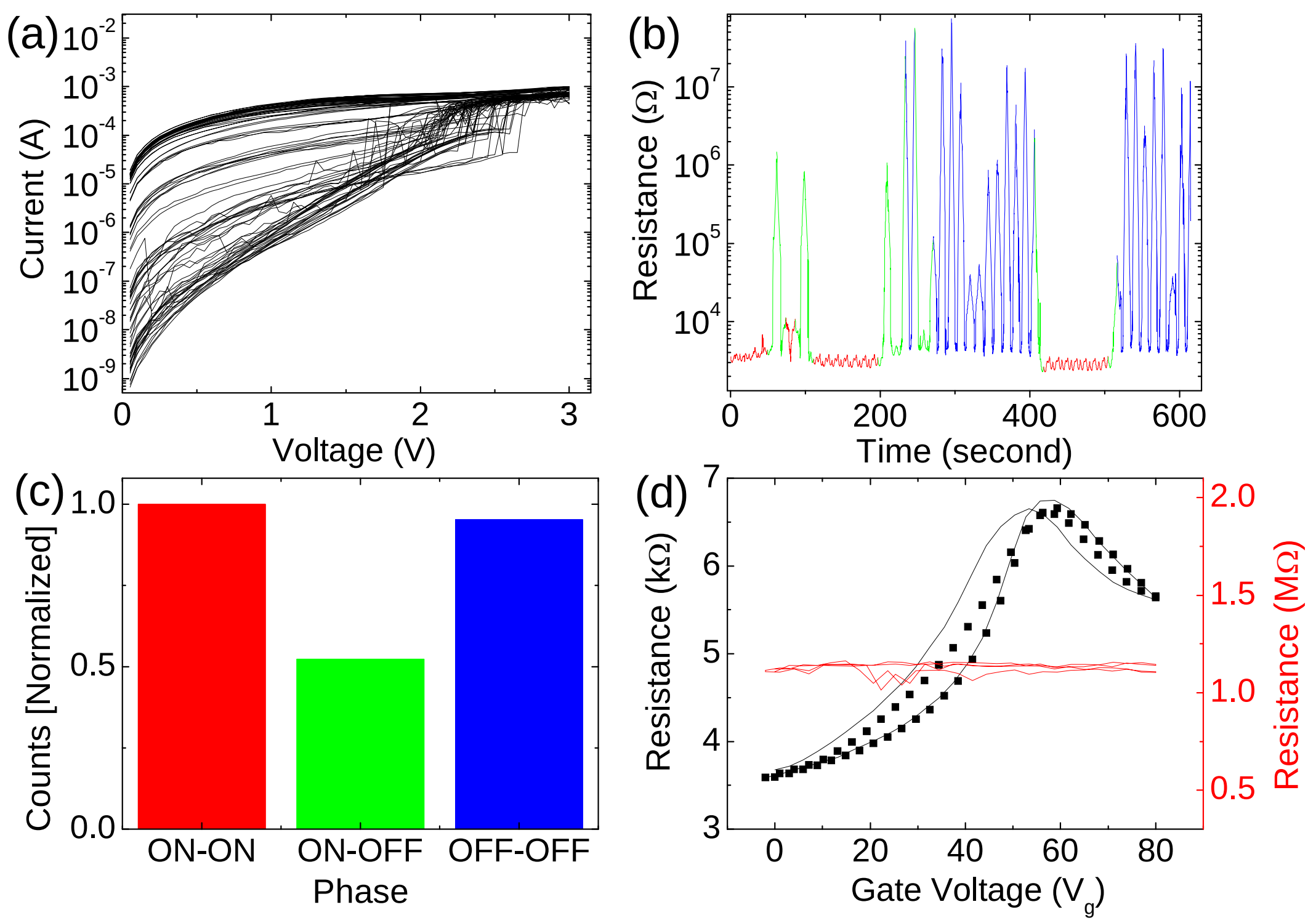

Figure 2.

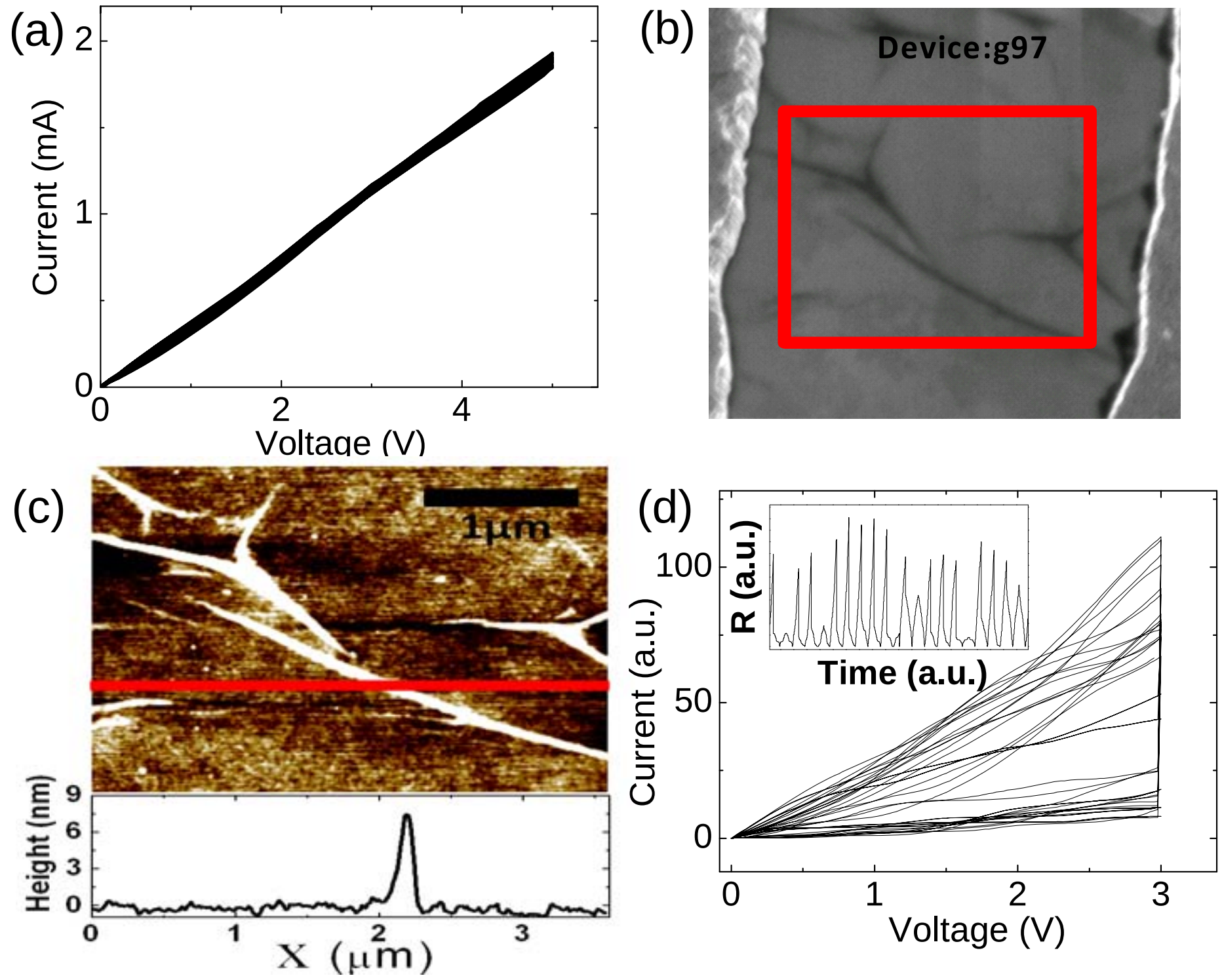

Figure 3.

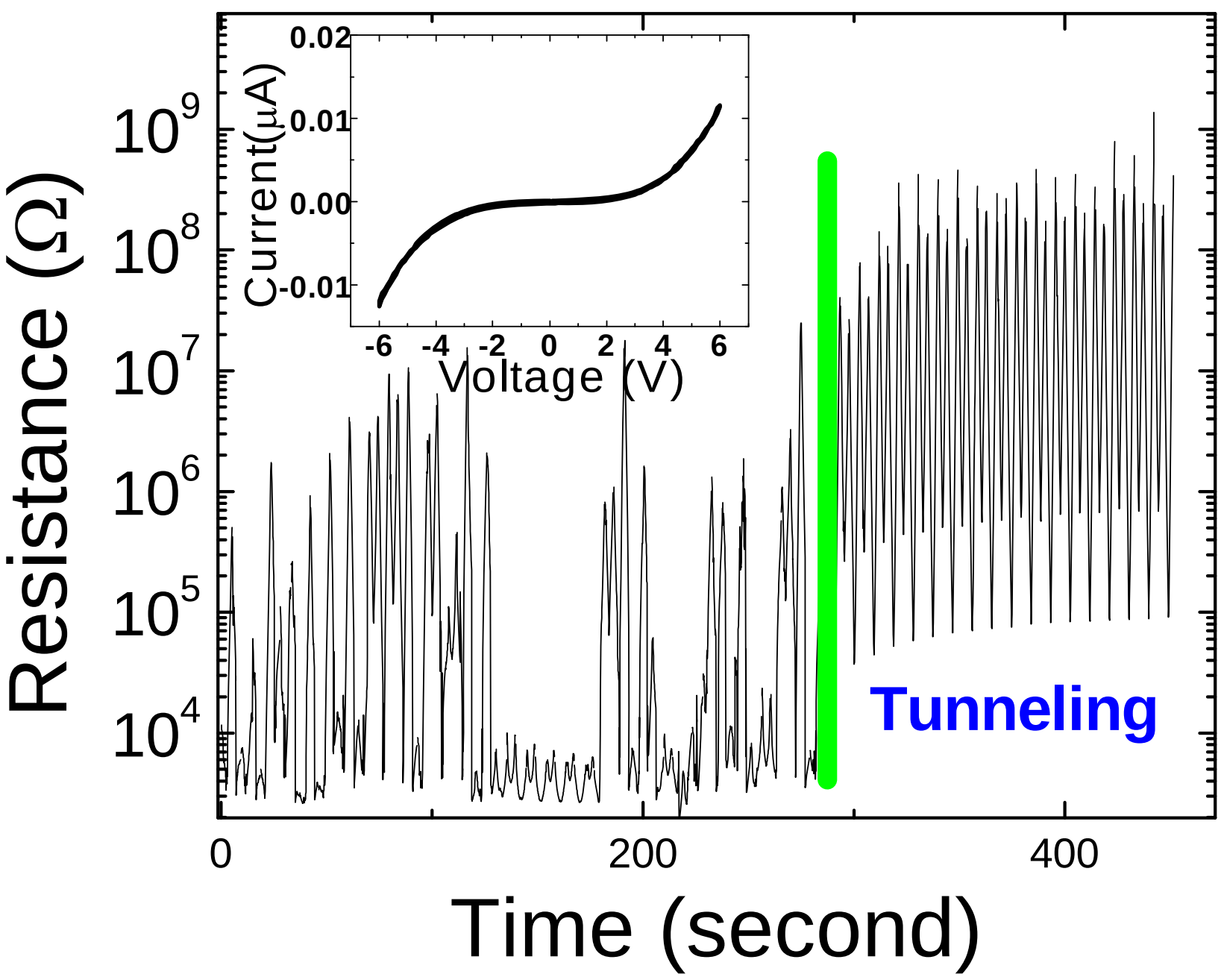

Figure 4.